# Generating Clustered Journal Maps:

# An Automated System for Hierarchical Classification

Loet Leydesdorff,*[a] Lutz Bornmann,[b] and Caroline S. Wagner [c]

**Abstract**

Journal maps and classifications for 11,359 journals listed in the combined Journal Citation Reports 2015 of the Science and Social Sciences Citation Indexes are provided at https://leydesdorff.github.io/journals/ and http://www.leydesdorff.net/jcr15. A routine using VOSviewer for integrating the journal mapping and their hierarchical clusterings is also made available. In this short communication, we provide background on the journal mapping/clustering and an explanation about and instructions for the routine. We compare journal maps for 2015 with those for 2014 and show the delineations among fields and subfields to be sensitive to fluctuations. Labels for fields and sub-fields are not provided by the routine, but an analyst can add them for pragmatic or intellectual reasons. The routine provides a means of testing one's assumptions against a baseline without claiming authority; clusters of related journals can be visualized to understand communities. The routine is generic and can be used for any 1-mode network.

**Keywords**: classification, visualization, journal, scientific field, citation, decomposition

[a] * corresponding author; Amsterdam School of Communication Research (ASCoR), University of Amsterdam PO Box 15793, 1001 NG Amsterdam, The Netherlands; loet@leydesdorff.net
[b] Division for Science and Innovation Studies, Administrative Headquarters of the Max Planck Society, Hofgartenstr. 8, 80539 Munich, Germany; bornmann@gv.mpg.de
[c] John Glenn College of Public Affairs, The Ohio State University, Columbus, Ohio, USA, 43210; wagner.911@osu.edu

## 1. Introduction

Scholarly journals have been and remain the primary organizers of scientific communication. The number of journals has increased over the centuries, at times showing exponential growth (Mabe, 2001; Mabe & Amin, 2003; Price, 1961, p. 166), but the journal form has remained remarkably stable in the social life of science. The intellectual development of the sciences and their organization, as well as growth of new specialties and disciplines, is organized, validated, and retained in scholarly journals. Ware & Mabe (2015) estimated that there were 28,100 peer-reviewed journals published in English in 2015; in the Web of Science (WoS) in that year, 11,365 journals were indexed. The source journals can be expected to account for more than 90 percent of citations because of the skew in the distributions (Garfield, 1971; Seglen, 1992).

Specialties and new fields develop at a level above individual journals. Since journals relate to one another through citations and references (Price, 1965), perhaps the best way to identify networked communities is through the cross-referencing of these already aggregated citations and references into algorithmically significant clusters (Leydesdorff, 1987; Tijssen *et al.*, 1987). This article describes a method of visualizing journal-to-journal connections to create 'macro-epistemics' (Knorr-Cetina, 2007). New developments with validity can be expected to form new journals and journal clusters (van den Besselaar & Leydesdorff, 1996).

The classification of journals into disciplines is complicated by the many venues where one finds results. Multidisciplinary journals such as *Science* and *Nature* play important roles in scientific communication, especially in calling attention to advances in knowledge. More recently, open-access journals (e.g., *PLoS ONE*) have emerged which deliberately ignore disciplinary boundaries and thus tend to disturb the classification of journals. Some scholars suggest that the journal form may diminish in use in favor of archives and repositories (Harnad, 2001), although the majority of scholars view journals as increasingly important (e.g., Marbán, 1999). As Lavoie *et al*. (2014) detail, "the transition from print to a digital, networked environment likely means that decision-making around the scholarly record will have to become more consciously coordinated."

Journals are classified into disciplinary groups by indexing services; the classifications serve a number of purposes. First, classification serves to facilitate the process of search and retrieval. Secondly, bibliometric evaluations use journal classifications to normalize citation scores (Moed, De Bruin, & Van Leeuwen, 1995; Schubert & Braun, 1986; Schubert, Glänzel, & Braun, 1986). For pragmatic reasons, it has been considered "best practice" in evaluation studies to use the WoS Subject Categories (WCs) [1] for the operationalization of fields of science even though these categories do not represent homogeneous sets (Leydesdorff & Bornmann, 2016). They are attributed to journals by manual indexing and have been elaborated incrementally for more than forty years by the providers of the database (Bensman & Leydesdorff, 2009; Pudovkin & Garfield, 2002, p. 1113). Journals can be attributed to more than one WC.

Beyond journal names and identity through sponsorship (e.g., by learned societies), articles can be classified in terms of co-citation, bibliographic coupling, or direct citation relations (Klavans & Boyack, 2016, in press). Clustering the database at the level of papers, however, requires access to large computing capacity and to entire copies of Scopus (Boyack *et al*., 2011) or the

[1] Before the introduction of WoS v.5 in 2009, the categories were referred to as ISI Subject Categories.

WoS (Waltman & van Eck, 2012). The problem of the validity of the delineations remains. As Schubert, Glänzel, & Braun (1989, at p. 7) have noted, "the field/subfield classification of papers is a neuralgic point of all kind of scientometric evaluations."

Aware of the constraints of using WCs for evaluation purposes, Glänzel & Schubert (2003) developed a new journal classification system based on a pragmatic weighting of the results of algorithmic clustering of journals in terms of citation patterns against expert judgment. The Centre for Research and Development Monitoring ECOOM of the Catholic University of Leuven (Belgium) uses this classification system for evaluations. In the meantime, fast decomposition algorithms have been developed that can be used for classifications. Klavans & Boyack (in press, at p. 12, Table 3) list seven journal-based partitions of Scopus data currently in use.

Rafols and Leydesdorff (2009) compared (*i*) the WCs and (*ii*) Glänzel & Schubert's (2003) alternative classification with two algorithmically generated ones: (*iii*) Newman & Girvan's (2004) algorithm applied to the matrix of 7,611 citing journals contained in the Journal Citation Reports (JCR) 2006; and (*iv*) a random-walk based algorithm submitted by Rosvall & Bergstrom (2008) that had been applied to 6,128 journals in the JCR 2004. The concordance between the four classifications was modest: in the 40-60% range (Rafols & Leydesdorff, 2009, Table 3, at p. 1828). This conclusion agrees with Boyack's estimate of 50% correct classifications for the WCs (Boyack, *personal communication,* 14 September 2008). However, most of the miscategorised journals appear to occur in areas within the close vicinity of categories indicated by the other classifications. In other words, the various decompositions are roughly consistent with each other, but imprecise. Despite the low correspondence, maps based on the different classifications can be rather similar (Leydesdorff & Rafols, 2009; Klavans & Boyack, 2009).

In summary, there are no unique or universally valid classifications of journals. Two runs of the same algorithmic decomposition may not provide the same results. Most algorithms begin by drawing a random number using the computer clock. However, Leydesdorff, Bornmann, & Zhou (2016, at p. 907) noted that VOSviewer—visualization software developed by CWTS and available free for download at http://www.vosviewer.com—can generate quasi-deterministic classifications when the seed number of the randomizer is set equal to a constant (the default is zero). Using this option, the decomposition can pragmatically be combined with visualizations in a hierarchical classification by using the output of each decomposition recursively as input to the further decomposition at a next-lower level (Waltman, van Eck, & Noyons, 2010). One begins at the top-level of the complete matrix and then extracts the clusters one by one; this process can be automated in a recursive loop if an option were added to VOSviewer for writing the output files to disk when running the program from the command line (Nees Jan van Eck, *personal communication*, 3 and 16 May 2016).

The most recent version 1.6.5 of VOSviewer (dated September 28, 2016), among other things, enables the user to run VOSviewer in a batch job from the command line. In this short communication, we report on generating such an automatic classification and visualization of the JCR-2015 data. The resulting classifications, visualizations, and routines are available at https://leydesdorff.github.io/journals/ and http://www.leydesdorff.net/jcr15 . The website provides input files for journal maps for the more than 11,000 journals contained in the JCR-2015, at the various levels discussed above.

Although developed for JCR-data, the routines are formulated so that any 1-mode matrix can be decomposed similarly in terms of mappings using VOSviewer. Note that one can also export the clusters in the Pajek format so that the files can be used for other visualizations such as in Gephi.

## 2. Data and Methods

### *2.1. Data*

Using dedicated software, the JCR 2015 data for the Science and Social Sciences Citation Indexes was first organized into a matrix of citing versus cited journals. All 11,365 journals are included so that the matrix is 1-mode, albeit asymmetrical. VOSviewer symmetrizes the asymmetrical matrix internally by summing the cells ($i,j$) and ($j,i$). Six journals are not connected (*Avian Res, EDN, Neuroforum, Austrian Hist Yearb, Curric Matters,* and *Policy Rev*), and are therefore excluded from the analysis. Thus, we work with (11,365 – 6 =) 11,359 journals. The results can be compared with results based on JCR-2014 data elaborated for a single branch in Leydesdorff, Bornmann, & Zhou (2016).[2] Table 1 provides descriptive statistics and network characteristics for the large components in 2015 and 2014. The intersection between these two years (using identical journal names) contains 11,009 journals.

[2] In 2014, the following six journals were not connected: *Edn, Argos-Venezuela, Balt J Econ, Curric Matters, Econtent,* and *Restaurator*.

**Table 1**: Network characteristics of the largest components of the matrix based on JCR 2015, compared with JCR 2014.

| | *JCR 2014* (a) | *JCR 2015* (b) | (c) |
|---|---|---|---|
| **N of journals (nodes)** | 11,143 | 11,359 | +1.9% |
| **Links** | 2,699,210 (10,829 loops) | 2,848,736 (11,049 loops) | +5.5% |
| **Total citations** | 40,787,243[3] | 43,010,234 | +5.5% |
| **Density** | 0.022 | 0.022 | 0 |
| **Average (total) degree** | 484.677 | 501.582 | +3.5% |
| **Cluster coefficient** | 0.220 | 0.220 | 0 |

Table 1 shows that the network increases more in terms of links than nodes. However, the density and the clustering coefficients did not change.

*2.2. Methods*

The routine (called "decomp.exe")[4] presumes an input file named "level0.net" containing the 1-mode network file saved in the .net format of Pajek. Decomp.exe begins with running the following statement from the command line:

> "C:\vosviewer\vosviewer -pajek_network C:\temp\level0.net -save_map C:\temp\m0.txt -save_network C:\temp\n0.txt -run_layout -run_clustering -repulsion 0 -min_cluster_size 2 -merge_small_clusters true"

VOSviewer is to be installed in the folder C:\vosviewer and one operates in the folder C:\temp . The "minimum cluster size" is set to "two" in order to suppress isolates; repulsion is set to "zero" to optimize the visualizations.

[3] Loops (that is, journal self-citations) were removed.
[4] Available at http://www.leydesdorff.net/jcr15/program.htm .

The initial output is written to the files m0.txt for the map and n0.txt for the network (at level 1), respectively. The file m0.txt contains the clustering that is used by the routine for generating an input file for each of the clusters. This next round generates output files m1.txt, m2.txt, etc., as map files of VOSviewer which contain the information for drawing maps at the next-lower level (level 2). In a next round, each of these files is further decomposed into m1_1.txt, m1_2.txt, etc. (level 3). The tree can be found at http://www.leydesdorff.net/jcr15/tree.htm. The two levels are attributed to individual journals at http://www.leydesdorff.net/jcr15/index.htm. Finally, all level-3 files are run in VOSviewer in order to generate the classification at level 4. This classification is attributed to each journal as a hyperlink at http://www.leydesdorff.net/jcr15: by clicking on a journal name, one webstarts VOSviewer to generate a map of the citation environment of this journal at level 4. The user can save this map for further decomposition (at level 5; see below).

## 3. Results

### *3.1. The global map based on JCR 2015 data*

Figure 1 provides the global map based on 2015 data. One can compare this map, for example, with the 2014 map (Leydesdorff, Bornmann, & Zhou, 2016, at p. 906);[5] the procedures for producing these two maps were virtually identical. However, the resulting delineations are notably different (Table 2; cf. Leydesdorff *et al.*, 2016a, Table 4, at p. 907).

[5] This map can be web-started at http://www.vosviewer.com/vosviewer.php?map=http://www.leydesdorff.net/journals14/jcr14.txt&cluster_colors=http://www.leydesdorff.net/journals14/colors14.txt&label_size_variation=0.3&zoom_level=1&scale=0.9

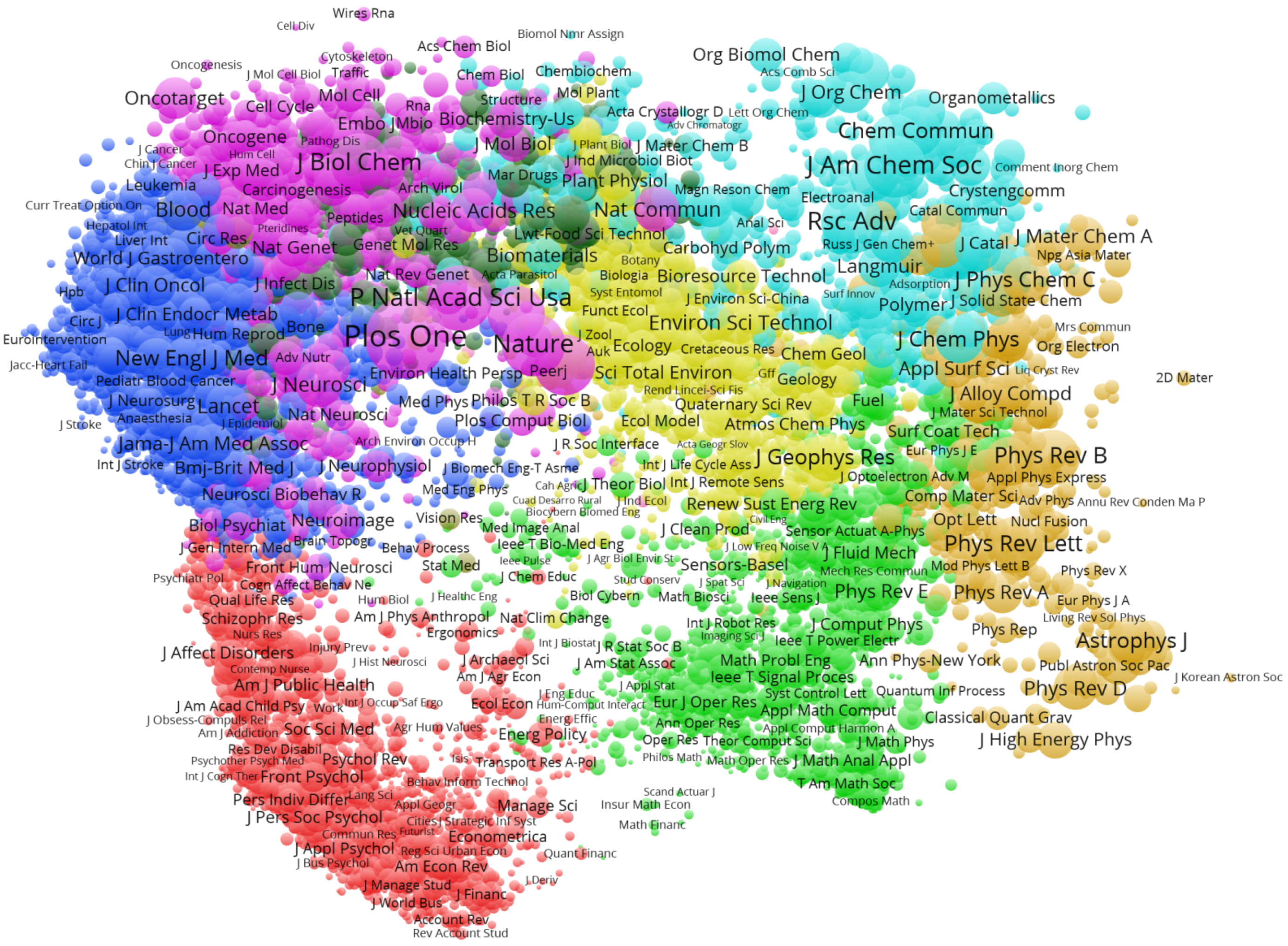

**Figure 1**: Ten clusters of 11,359 journals (largest component of the JCR matrix) based on 2015 data; VOSviewer used for classification and visualization. This map can be web-started at http://tinyurl.com/jmrwp64 or http://www.vosviewer.com/vosviewer.php?map=http://www.leydesdorff.net/jcr15/m0.txt&label_size_variation=0.3&zoom_level=1.5&cluster_colors=http://www.leydesdorff.net/jcr15/colors.txt&scale=0.9

**Table 2**: Fields distinguished at the top level of JCR 2015 and 2014.

| | 2014 | | 2015 | | |
|---|---|---|---|---|---|
| | *Field-designation* | *N* | *Field-designation* | *N* | *Color in Fig. 1* |
| 1 | Social Sciences | 3,131 | Social Sciences | 3,274 | red |
| 2 | Medicine | 1,943 | Computer Science | 2,003 | green |
| 3 | Computer Science | 1,939 | Medicine | 1,965 | blue |
| 4 | Environmental Sci | 1,911 | Environmental Sci | 1,595 | yellow |
| 5 | Chemistry | 684 | Biomedical | 784 | lila |
| 6 | Biomedical | 672 | Chemistry | 652 | light-blue |
| 7 | Physics | 462 | Bio-agricultural | 583 | dark-green |
| 8 | Neuro Sciences | 343 | Physics | 440 | orange |
| 9 | Ophthalmology | 56 | Ophthalmology | 57 | brown |
| 10 | | | Data analysis ("Big data") | 6 | pink |
| | | 11,141 | | 11,359 | |

Since the clustering is hierarchical, the extraction of different sets can sometimes become a trade-off among memberships of journals in different groups. For example, in 2014, as can be seen in Table 2, an eighth cluster of 343 neuroscience journals is distinguished. This same cluster is no longer visible in 2015; the same journals are split between a third cluster ("Medicine") and a fifth cluster ("Biomedical"). In 2015, however, cluster seven (dark green in Figure 1) groups 583 journals into a "bio-agricultural" cluster. The extraction of this seventh set (before the extraction of the neuroscience group as the eighth set) changes the path of the decomposition so that a different sub-optimum is reached. Note that this different branching can be caused by relatively small differences in the data.

The program does not provide the disciplinary designations; labels can be added (subjectively) to the algorithmic artifacts by the analyst depending of the objectives of the study. As shown in Table 2, a tenth field of only six journals is identified in 2015. We have labeled this cluster "data analysis" based upon the titles of the six journals (Table 3) and their combined citation environment (Figure 2).[6] While the larger environment of the cluster (indicated as "#Big Data-US") shows biomedicine, environmental sciences, chemistry, etc., the *J Am Stat Assoc*, a core journal in statistics, is mapped in close proximity to the cluster as is *Commun ACM*, a leading journal in computer science.

**Table 3:** Six journals in cluster 10

| |
|---|
| Big Data-US |
| Environ Sci Tech Let |
| Environ Sci-Nano |
| J Ind Ecol |
| Microbiome |
| Omics |

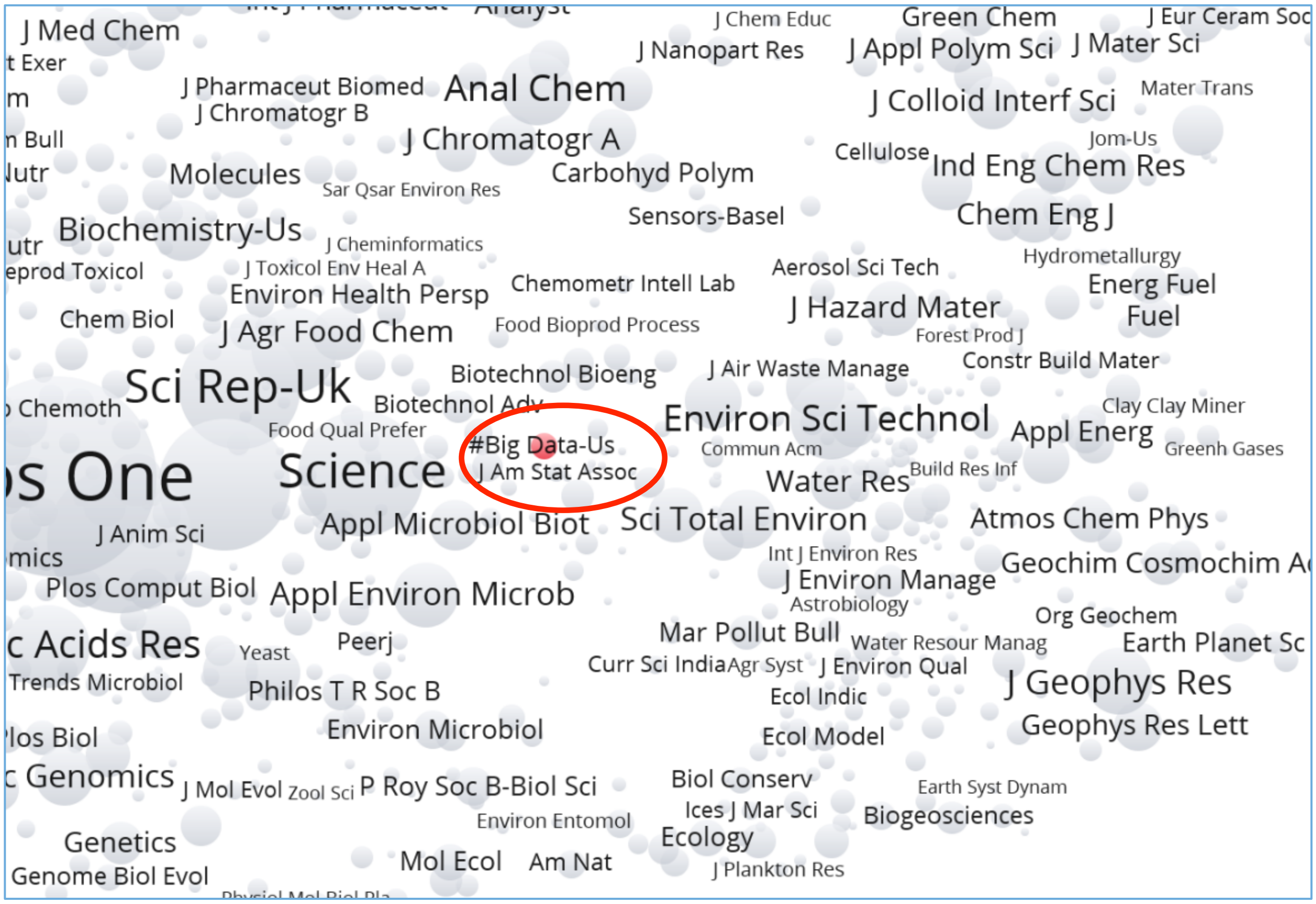

**Figure 2**: Zoom of the *k*=1 (citation) environment of cluster 10 (*N* = 1236).

[6] This environment was generated by shrinking the tenth partition of six journals into a macro-journal, of which the *k*=1 neighbourhood can be determined in Pajek. This direct citation environment (citing and cited) contains 1236 journals.

The classification in 2015 (ten clusters) can be compared with the one in 2014 (nine clusters) for the 11,009 journals that are included in the JCR versions of both years. The chi-square of the cross-tabulation is highly significant ($p<.001$), but Cramer's $V$—a measure of the chi-square which varies between zero and one—is only 0.82. Ten percent of the journals are differently classified between 2014 and 2015. Although the clusters are reproducible within each year, the clustering is, in our opinion, not sufficiently reliable for comparisons across years. As noted, relatively small changes in numbers of citations can affect the order of the extractions in a hierarchical decomposition.

### *3.2. The social-sciences cluster*

In both 2014 (with 3131) and 2015 (with 3274), cluster 1—representing the social sciences--is the largest group. This is not a homogenous cluster; but the citation patterns in the social sciences are statistically so different from those in natural sciences and engineering that they are sorted separately in the first pass of the decomposition (at level 1). Figure 3 provides the decomposition of this cluster at level 2; the information is summarized in Table 4 and compared with the corresponding table for 2014.

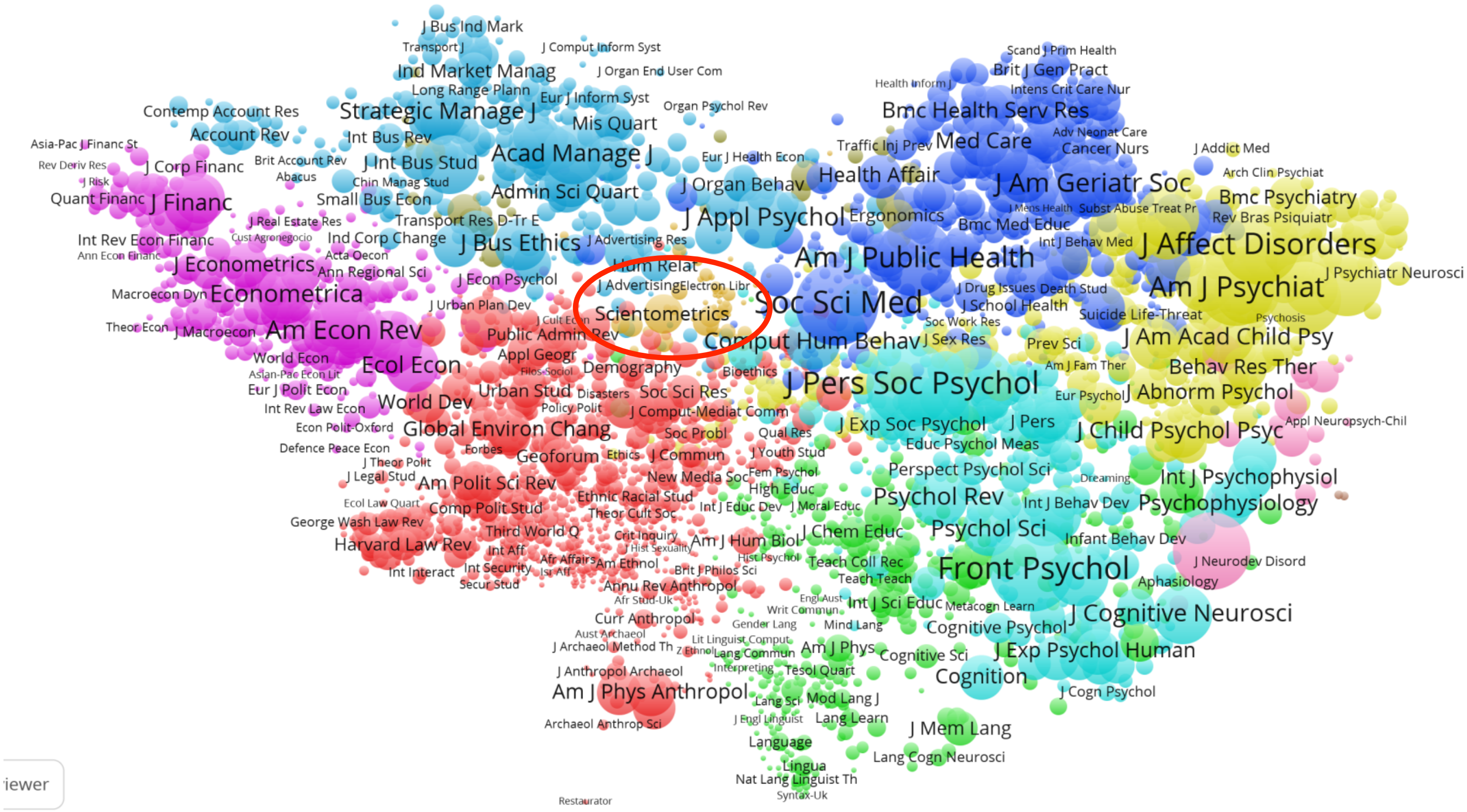

**Figure 3:** Decomposition of the social-sciences cluster based on 2015 data. This file can be web-started at http://tinyurl.com/gl6unrc or http://www.vosviewer.com/vosviewer.php?map=http://www.leydesdorff.net/jcr15/m1.txt&label_size_variation=0.35&scale=0.9

**Table 4**: Decomposition of the social-sciences cluster in 2014 and 2015.[7]

| | 2014 | N | 2015 | N | Color in Fig. 2 |
|---|---|---|---|---|---|
| 1 | Discipline-oriented social science | 1,008 | Discipline-oriented social sciences | 1069 | Red |
| 2 | Application-oriented social science | 385 | Language and education | 459 | Green |
| 3 | Health | 345 | Health | 412 | Dark blue |
| 4 | Economics | 335 | Psychiatry | 329 | Light yellow |
| 5 | Mental Health | 267 | Economics | 317 | Dark purple |
| 6 | Administration | 255 | Psychology | 287 | Light blue |
| 7 | Language | 188 | Management Science | 278 | Blue |
| 8 | Psychology | 146 | Library & Information Science | 62 | Light brown |
| 9 | Law | 117 | Transport | 38 | Dark brown |
| 10 | Library & Information Science | 52 | Neuropsychology | 21 | Light purple |
| 11 | Transport | 33 | (Hypnosis) | 2 | Dark yellow |
| | Sum | 3,131 | Sum | 3,274 | |

[7] The overlap of journals (with the same name) between 2014 and 2015 contains 3,073 journals; Cramer's $V$ between the two classifications is 0.78.

Note that a light-brown cluster with 62 “library and information science” journals can be found in the middle of Figure 1 (indicated most clearly by the journal title “*Scientometrics*” circled).

*3.3. Library and information sciences*

Figure 4 provides the map for the 62 LIS journals distinguished as a cluster in 2015. As in 2014, the group of journals related to management information is not included in the LIS set; but differently from 2014, a (green-colored) group of statistics and methods journals is now included (at the right side of the figure). One can generate this map by clicking on one of these journals at http://www.leydesdorff.net/jcr15.

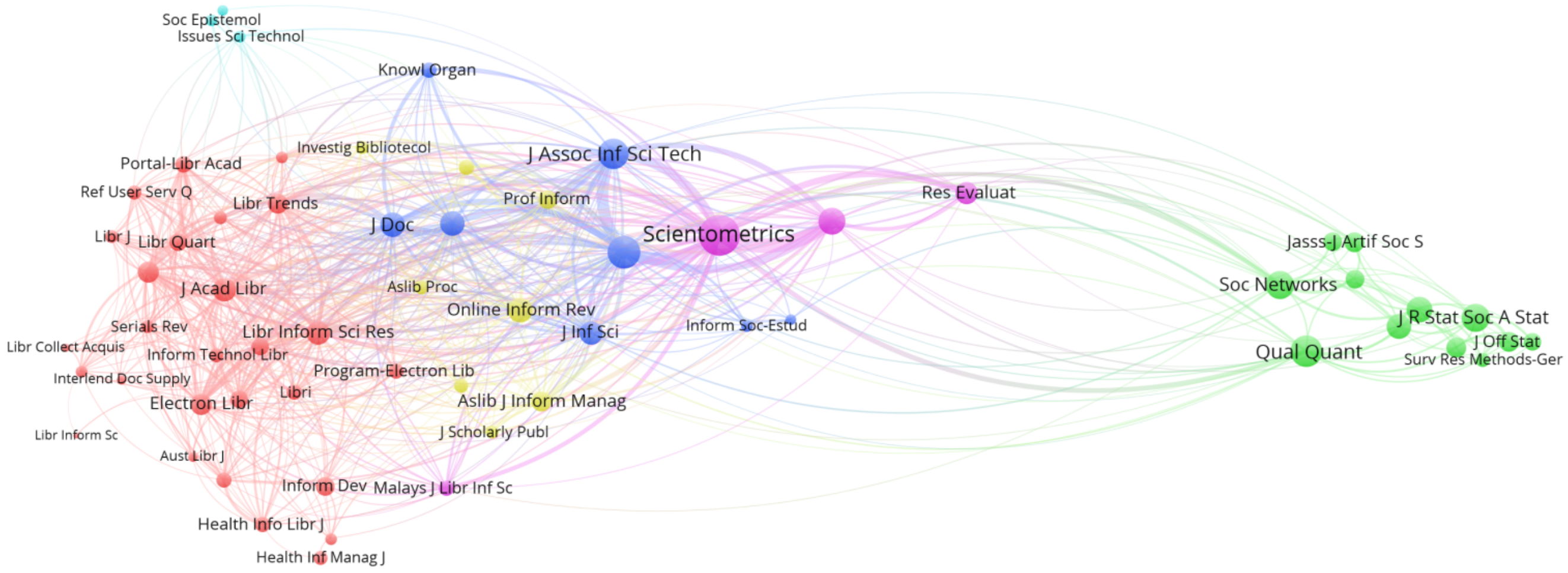

**Figure 4**: Map with sixty-two LIS journals in 2015. This file can be web-started at http://tinyurl.com/gvyafak or http://www.vosviewer.com/vosviewer.php?map=http://www.leydesdorff.net/jcr15/m1_8.txt&network=http://www.leydesdorff.net/jcr15/n1_8.txt&label_size_variation=0.25&scale=1.25&colored_lines&curved_lines&n_lines=10000

In Appendix 1, the two sets (for 2015 and 2014) are compared with the WC “information science and library science”. Forty-three journals co-occur in all three lists; 49 co-occur in two of the three lists. The WC also includes 37 journals that belong mostly to the cluster of management-information-science journals (Leydesdorff & Bornmann, 2016).

As noted, the JCR-2015 set includes 12 journals that belong to a "statistics and methods" cluster. Journals such as *Social Networks* are cited both in "information science" and in other fields such as "business & management" or "organization studies" (Leydesdorff *et al*., 2008). In 2014, for example, this journal is grouped with the *J Artif Soc S* in a cluster of 143 sociology journals, whereas *Qual Quant* is grouped among 335 economic journals. However, one is dis-advised to draw far-reaching conclusions on the basis of changes among two subsequent years (Leydesdorff & de Nooy, in press).

Further decomposition of the LIS set leads to six clusters which vary from three to 27 journals (Table 4). In other words, the relatively homogenous modules of the (social) sciences may be rather fine-grained. In our opinion, this leads to the question of whether one can use these clusters to normalize citation behavior above the level of individual journals. The delineations among "fields" and "subfields" (in terms of citation patterns) seem sensitive to weak fluctuations that might, from another perspective, be considered as noise (Leydesdorff, 2006).

**Table 4**: Decomposition of LIS cluster 2015 (62 journals) at level 5

| | |
|---|---|
| Library science | 27 |
| Methodology | 12 |
| Information science | 8 |
| Publishing | 8 |
| Bibliometrics | 4 |
| Meta-issues | 3 |
| | 62 |

Patterns may be affected by specific events. For example, the publication of one or two special issues on the border between two specialisms may change the pattern and provide the impression of emerging new developments. From this perspective, one can question the suggestion made above that a new set of six journals were labeled as "data analysis" or "big data". This may be an over-interpretation on our side, influenced by the hype around this topic. Moreover, many articles about "big data" appear in journals other than the six journals listed in Table 3.

## 4. Discussion and Conclusions

The matrix of aggregated journal-journal citation relations represents a complex system of scientific communication that is both hierarchically layered and functionally differentiated in terms of scientific specialties and fields. Such a system cannot be decomposed unambiguously (Simon, 1973). The clusters can be related in other (e.g., methodological versus theoretical) dimensions; densities of communication in subsets can vary significantly. Referencing behavior norms may differ across fields. For example, an article in a biomedical specialty may contain more than forty references, while in other fields, such as mathematics, fewer than ten references is more common (Garfield, 1979; Moed, 2010). However, what is measured as "differences in citation behavior among fields" can also be an artifact of the different degrees of coverage of the field-specific literature by the database (Marx & Bornmann, 2015). Epistemically, references may function at research fronts to position the *citing* papers or acknowledge intellectual debt and/or credit to previous (that is, *cited*) publications (Leydesdorff, Bornmann, Comins, & Milojević, 2016). Bodies of specialist literature may interact in next-order—i.e., more generalist—layers carried by quality journals such as *Science* and *Nature*.

This complex interweaving of different dynamics is further complicated because all relevant distributions are heavily skewed (Seglen, 1992). Weak ties in one context can be strong ties from another perspective (Granovetter, 1973). As we have seen above, hierarchical decomposition follows a path downward so that the results are path-dependent and may lead to different sub-optima. There is no objective yardstick to inform us how much better one representation is when compared with another (cf. Klavans & Boyack, 2016, in press).

In addition to the statistical quality of the distinctions, the groupings have to be labeled manually; this adds a subjective dimension of flexible interpretations with different meanings, since the labels are not provided by the decomposition itself. The labels are added by an analyst who, as a user of the system, may wish to mix pragmatic with intellectual considerations. *Ex ante*, one representation is as legitimate as another and no methodological prescription can be formulated.

Within this context of uncertainty and complexity, the proposed routine provides a means for testing one's assumptions without claiming authority; but with the advantage of reproducibility and the possibility of rich visualizations. The algorithm is semantically neutral: the routine will work on any 1-mode matrix and provide a purely algorithmic decomposition of the system into lower-level units in a series of layers. The advantages of using this decomposition and the quality of the visualizations will have to show their usefulness in bibliometric practices. The results may raise further questions and thus help to shape research ideas and agendas.

## Acknowledgements

We thank Nees Jan van Eck and Ludo Waltman for the adaptation of VOSviewer and for comments and suggestions. We thank also Kevin Boyack for suggestions. We are grateful to Thomson Reuters for providing us with JCR data.

## References

Bensman, S. J., & Leydesdorff, L. (2009). Definition and Identification of Journals as Bibliographic and Subject Entities: Librarianship vs. ISI Journal Citation Reports (JCR) Methods and their Effect on Citation Measures. *Journal of the American Society for Information Science and Technology, 60*(6), 1097-1117.

Boyack, K. W., Newman, D., Duhon, R. J., Klavans, R., Patek, M., Biberstine, J. R., . . . Börner, K. (2011). Clustering more than two million biomedical publications: Comparing the accuracies of nine text-based similarity approaches. *PLoS ONE, 6*(3), e18029.

Garfield, E. (1971). The mystery of the transposed journal lists—wherein Bradford's Law of Scattering is generalized according to Garfield's Law of Concentration. *Current Contents, 3*(33), 5–6.

Garfield, E. (1979). Is citation analysis a legitimate evaluation tool? *Scientometrics, 1*(4), 359-375.

Glänzel, W., & Schubert, A. (2003). A new classification scheme of science fields and subfields designed for scientometric evaluation purposes. *Scientometrics, 56*(3), 357-367.

Granovetter, M. S. (1973). The strength of weak ties. *American Journal of Sociology, 78*(6), 1360-1380.

Harnad, S. (2001). Why I think research access, impact and assessment are linked. *Times Higher Education Supplement, 1487*, 16.

Klavans, R., & Boyack, K. (2009). Towards a Consensus Map of Science. *Journal of the American Society for Information Science and Technology, 60*(3), 455-476.

Klavans, R., & Boyack, K. W. (2016, early view). Which type of citation analysis generates the most accurate taxonomy of scientific and technical knowledge? *Journal of the Association for Information Science and Technology*. doi: 10.1002/asi.23734

Knorr-Cetina, K. (2007). Culture in global knowledge societies: Knowledge cultures and epistemic cultures. *Interdisciplinary science reviews*, *32*(4), 361-375.

Lavoie, B., Childress, E. Erway, R. Faniel, I., Malpas, C., Schaffner, J., Van der Werf, T. (2014). *The Evolving Scholarly Record*. Dublin, Ohio: OCLC Research. http://www.oclc.org/research/publications/library/2014/oclcresearch-evolvingscholarly-record-2014.pdf.

Leydesdorff, L. (1987). Various methods for the mapping of science. *Scientometrics, 11*(5), 295-324.

Leydesdorff, L. (2006). Can Scientific Journals be Classified in Terms of Aggregated Journal-Journal Citation Relations using the Journal Citation Reports? *Journal of the American Society for Information Science & Technology, 57*(5), 601-613.

Leydesdorff, L., & Bornmann, L. (2016). The Operationalization of "Fields" as WoS Subject Categories (WCs) in Evaluative Bibliometrics: The cases of "Library and Information Science" and "Science & Technology Studies". *Journal of the Association for Information Science and Technology, 67*(3), 707-714.

Leydesdorff, L., Bornmann, L., Comins, J., & Milojević, S. (2016). Citations: Indicators of Quality? The Impact Fallacy. *Frontiers in Research Metrics and Analytics, 1*(Article 1). doi: 10.3389/frma.2016.00001

Leydesdorff, L., Bornmann, L., & Zhou, P. (2016). Construction of a pragmatic base line for journal classifications and maps based on aggregated journal-journal citation relations. *Journal of Informetrics, 10*(4), 902-918.

Leydesdorff, L., & de Nooy, W. (in press). Can "Hot Spots" in the Sciences Be Mapped Using the Dynamics of Aggregated Journal-Journal Citation Relations? *Journal of the Association for Information Science and Technology*, at http://arxiv.org/abs/1502.00229.

Leydesdorff, L., & Rafols, I. (2009). A Global Map of Science Based on the ISI Subject Categories. *Journal of the American Society for Information Science and Technology, 60*(2), 348-362.

Leydesdorff, L., Schank, T., Scharnhorst, A., & De Nooy, W. (2008). Animating the Development of *Social Networks* over Time using a Dynamic Extension of Multidimensional Scaling. *El Profesional de la Información, 17*(6), 611-626.

Mabe, M. (2003). The growth and number of journals. *Serials*, *16*(2), 191-197.

Mabe, M., & Amin, M. (2001). Growth dynamics of scholarly and scientific journals. *Scientometrics*, *51*(1), 147-162.

Marbán, E. (1999). Inaugural Editorial: A New Era for Circulation Research. *Circulation Research*, *85*(1), 1-3.

Marx, W., & Bornmann, L. (2015). On the causes of subject-specific citation rates in Web of Science. *Scientometrics, 102*(2), 1823-1827.

Moed, H. F. (2010). Measuring contextual citation impact of scientific journals. *Journal of Informetrics, 4*(3), 265-277.

Moed, H. F., De Bruin, R. E., & Van Leeuwen, T. N. (1995). New bibliometric tools for the assessment of national research performance: Database description, overview of indicators and first applications. *Scientometrics, 33*(3), 381-422.

Newman, M. E., & Girvan, M. (2004). Finding and evaluating community structure in networks. *Physical Review E, 69*(2), 026113.

Price, D. J. de Solla (1961). *Science Since Babylon*. New Haven: Yale University Press.

Price, D.J. de Solla (1965). Networks of scientific papers. *Science*,*149*(3683), 510-515.

Pudovkin, A. I., & Garfield, E. (2002). Algorithmic procedure for finding semantically related journals. *Journal of the American Society for Information Science and Technology, 53*(13), 1113-1119.

Rafols, I., & Leydesdorff, L. (2009). Content-based and Algorithmic Classifications of Journals: Perspectives on the Dynamics of Scientific Communication and Indexer Effects. *Journal of the American Society for Information Science and Technology, 60*(9), 1823-1835.

Rosvall, M., & Bergstrom, C. T. (2008). Maps of random walks on complex networks reveal community structure. *Proceedings of the National Academy of Sciences, 105*(4), 1118-1123.

Schubert, A., & Braun, T. (1986). Relative indicators and relational charts for comparative assessment of publication output and citation impact. *Scientometrics, 9*(5), 281-291.

Schubert, A., Glänzel, W., & Braun, T. (1986). Relative indicators of publication output and citation impact of european physics research, 1978–1980. *Czechoslovak Journal of Physics, 36*(1), 126-129.

Schubert, A., Glänzel, W., & Braun, T. (1989). World Flash on Basic Research. Scientometric Datafiles. A Comprehensive Set of Indicators on 2649 Journals and 96 Countries in All Major Science Fields and Subfields 1981-1985. *Scientometrics 16*(1-6), 3-478.

Seglen, P. O. (1992). The Skewness of Science. *Journal of the American Society for Information Science, 43*(9), 628-638.

Simon, H. A. (1973). The Organization of Complex Systems. In H. H. Pattee (Ed.), *Hierarchy Theory: The Challenge of Complex Systems* (pp. 1-27). New York: George Braziller Inc.

Small, H. (1973). Co-citation in the scientific literature: A new measure of the relationship between two documents. *Journal of the American Society for information Science*, *24*(4), 265-269.

Small, H., & Griffith, B. C. (1974). The structure of scientific literatures I: Identifying and graphing specialties. *Science studies*, *4*(1), 17-40.

Tijssen, R., de Leeuw, J., & van Raan, A. F. J. (1987). Quasi-Correspondence Analysis on Square Scientometric Transaction Matrices. *Scientometrics 11*(5-6), 347-361.

Van den Besselaar, P., & Leydesdorff, L. (1996). Mapping Change in Scientific Specialties: A Scientometric Reconstruction of the Development of Artificial Intelligence. *Journal of the American Society for Information Science, 47*(6), 415-436.

Waltman, L., & van Eck, N. J. (2012). A new methodology for constructing a publication-level classification system of science. *Journal of the American Society for Information Science and Technology, 63*(12), 2378-2392.

Waltman, L., van Eck, N. J., & Noyons, E. (2010). A unified approach to mapping and clustering of bibliometric networks. *Journal of Informetrics, 4*(4), 629-635.

Ware, M., & Mabe, M. (2015). The STM report: An overview of scientific and scholarly journal publishing. The Hague: International Association of Scientific, Technical and Medical Publishers. Retrieved from http://digitalcommons.unl.edu/scholcom/9/, October 10, 2016.

Appendix 1

| | | **86 journals in WoS WC “information & library science”** | |
|---|---|---|---|
| **62 LIS journals in 2015** | **52 LIS journals in 2014** | ***49 overlapping*** | ***37 additional*** |
| Afr J Libr Arch Info | Afr J Libr Arch Info | Afr J Libr Arch Info | Data Base Adv Inf Sy |
| Aslib J Inform Manag | Aslib J Inform Manag | | Econtent |
| Aslib Proc | Aslib Proc | Aslib Proc | Ethics Inf Technol |
| Aust Acad Res Libr | Aust Acad Res Libr | Aust Acad Res Libr | Eur J Inform Syst |
| Aust Libr J | Aust Libr J | Aust Libr J | Gov Inform Q |
| Can J Inform Lib Sci | Can J Inform Lib Sci | Can J Inform Lib Sci | Inform Manage-Amster |
| Coll Res Libr | Coll Res Libr | Coll Res Libr | Inform Organ-Uk |
| Comput Math Organ Th | | | Inform Soc |
| Electron Libr | Electron Libr | Electron Libr | Inform Syst J |
| Eng Stud | | | Inform Syst Res |
| Field Method | | | Inform Technol Dev |
| Health Inf Manag J | | | Inform Technol Manag |
| Health Info Libr J | Health Info Libr J | Health Info Libr J | Inform Technol Peopl |
| | Inf Tarsad | | Int J Comp-Supp Coll |
| | Inform Cult | Inform Cult | Int J Geogr Inf Sci |
| Inform Dev | Inform Dev | Inform Dev | Int J Inform Manage |
| | Inform Process Manag | Inform Process Manag | J Am Med Inform Assn |
| Inform Res | Inform Res | Inform Res | J Assoc Inf Syst |
| Inform Soc-Estud | Inform Soc-Estud | Inform Soc-Estud | J Comput-Mediat Comm |
| Inform Technol Libr | Inform Technol Libr | Inform Technol Libr | J Glob Inf Manag |
| Interlend Doc Supply | | Interlend Doc Supply | J Glob Inf Tech Man |
| Investig Bibliotecol | Investig Bibliotecol | Investig Bibliotecol | J Health Commun |
| Issues Sci Technol | Issues Sci Technol | | J Inf Technol |
| J Acad Libr | J Acad Libr | J Acad Libr | J Knowl Manag |
| J Am Soc Inf Sci Tec | J Am Soc Inf Sci Tec | J Am Soc Inf Sci Tec | J Manage Inform Syst |
| J Assoc Inf Sci Tech | J Assoc Inf Sci Tech | J Assoc Inf Sci Tech | J Organ End User Com |
| J Doc | J Doc | J Doc | J Strategic Inf Syst |
| J Inf Sci | J Inf Sci | J Inf Sci | Knowl Man Res Pract |
| J Informetr | J Informetr | J Informetr | Mis Q Exec |
| | J Legal Educ | | Mis Quart |
| J Libr Inf Sci | J Libr Inf Sci | J Libr Inf Sci | Restaurator |
| J Math Sociol | | | Rev Esp Doc Cient |
| J Med Libr Assoc | J Med Libr Assoc | J Med Libr Assoc | Scientist |
| J Off Stat | | | Soc Sci Comput Rev |
| J R Stat Soc A Stat | | | Soc Sci Inform |
| J Scholarly Publ | J Scholarly Publ | J Scholarly Publ | Telecommun Policy |
| Jasss-J Artif Soc S | | | Telemat Inform |
| Knowl Organ | Knowl Organ | Knowl Organ | |
| | Law Libr J | Law Libr J | |
| Learn Publ | Learn Publ | Learn Publ | |

| | | |
|---|---|---|
| Libr Collect Acquis | Libr Collect Acquis | Libr Collect Acquis |
| Libr Hi Tech | Libr Hi Tech | Libr Hi Tech |
| Libr Inform Sc | Libr Inform Sc | Libr Inform Sc |
| Libr Inform Sci Res | Libr Inform Sci Res | Libr Inform Sci Res |
| Libr J | Libr J | Libr J |
| Libr Quart | Libr Quart | Libr Quart |
| Libr Resour Tech Ser | Libr Resour Tech Ser | Libr Resour Tech Ser |
| Libr Trends | Libr Trends | Libr Trends |
| Libri | Libri | Libri |
| Malays J Libr Inf Sc | Malays J Libr Inf Sc | Malays J Libr Inf Sc |
| Online Inform Rev | Online Inform Rev | Online Inform Rev |
| Portal-Libr Acad | Portal-Libr Acad | Portal-Libr Acad |
| Prof Inform | Prof Inform | Prof Inform |
| Program-Electron Lib | Program-Electron Lib | Program-Electron Lib |
| Qual Quant | | |
| Ref User Serv Q | Ref User Serv Q | Ref User Serv Q |
| Res Evaluat | Res Evaluat | Res Evaluat |
| Rev Esp Doc Cient | Rev Esp Doc Cient | |
| Scientometrics | Scientometrics | Scientometrics |
| Serials Rev | Serials Rev | Serials Rev |
| Soc Epistemol | | |
| Soc Networks | | |
| Sociol Method Res | | |
| Sociol Methodol | | |
| Surv Methodol | | |
| Surv Res Methods-Ger | | |
| Transinformacao | Transinformacao | Transinformacao |
| | Z Bibl Bibl | Z Bibl Bibl |